\documentclass[12pt]{article}
 \usepackage{graphicx}
 \usepackage[cp1251]{inputenc}

\begin{document}
 \noindent {\footnotesize\it
   Astronomy Letters, 2022, Vol. 48, No 7, pp. 376--388}
 \newcommand{\dif}{\textrm{d}}

 \noindent
 \begin{tabular}{llllllllllllllllllllllllllllllllllllllllllllll}
 & & & & & & & & & & & & & & & & & & & & & & & & & & & & & & & & & & & & & &\\\hline\hline
 \end{tabular}

  \vskip 0.5cm
\centerline{\bf\large Parameters of the Galactic Spiral Density Wave from Masers with}
\centerline{\bf\large Parallax Errors Less Than 10\%}

   \bigskip
   \bigskip
  \centerline
 {V. V. Bobylev \footnote [1]{e-mail: vbobylev@gaoran.ru} and A. T. Bajkova}
   \bigskip

  \centerline{\small\it Pulkovo Astronomical Observatory, Russian Academy of Sciences,}

  \centerline{\small\it Pulkovskoe sh. 65, St. Petersburg, 196140 Russia}
 \bigskip
 \bigskip
 \bigskip

{\bf Abstract}---We have studied the kinematics of Galactic masers and radio stars with measured VLBI trigonometric parallaxes and proper motions. We have considered masers with relative trigonometric parallax errors less than 10\% and determined the Galactic rotation parameters from them. In particular, the linear rotation velocity of the Galaxy at the solar distance $R_0$ has been found to be $244.4\pm4.3$~km s$^{-1}$ (for the adopted $R_0=8.1\pm0.1$~kpc). We have performed a joint and separate spectral analysis of the
radial, residual tangential, and vertical maser velocities. For example, from the vertical maser velocities we have estimated the velocity perturbation amplitude $f_W=5.2\pm1.5$~km s$^{-1}$ with the wavelength $\lambda_W=2.6\pm0.7$~kpc, arguing for the influence of the spiral density wave on the vertical stellar velocities. Based on 104 masers within 3~kpc of the Sun, as a result of the joint solution, we have estimated the radial,
$f_R=6.7\pm1.1$~km s$^{-1}$, and tangential, $f_\theta=2.6\pm1.2$~km s$^{-1}$, velocity perturbations, the perturbation wavelength $\lambda=2.1\pm0.3$~kpc, and the Sun's phase in the Galactic spiral density wave $\chi_\odot=-148\pm15^\circ$. We have confirmed the presence of the Radcliffe wave in the spatial distribution of masers and radio stars belonging to the Local Arm.


 \subsection*{INTRODUCTION}
We know several classes of objects that are of paramount importance for studying the structure and kinematics of the Galaxy owing to the possibility to estimate the distances to them with an acceptable accuracy. These include, for example, star clusters, OB stars, or Cepheids. Applying VLBI to the problem of measuring the trigonometric parallaxes of
Galactic masers has made them first-class objects for studying the Galaxy. In this paper, only the masers associated with the youngest stars and protostars that are in active star-forming regions will be of interest to us.

At present, there are $\sim$200 trigonometric parallaxes of masers (Reid et al. 2019; Hirota et al. 2020) measured with a high astrometric accuracy. The random VLBI measurement errors for most of these sources are less than 0.020 milliarcseconds (mas). For example, we know the result of an essentially direct determination of the Galactocentric distance of
the Sun by this measuring method, $R_0.$ The case in point is the parallax measurement for the radio source Sgr B, $\pi=0.129\pm0.012$~mas, giving an estimate of $R_0=7.9^{+0.8}_{-0.7}$~kpc (Reid et al. 2009).

Quite a few studies devoted to analyzing the fundamental properties of the Galactic disk using masers have been performed. Note the papers by Honma et al. (2012), Sanna et al. (2017), Rastorguev et al. (2017), Xu et al. (2018), Reid et al. (2019), Hirota et al. (2020), and Bobylev et al. (2020), where the distance $R_0,$ the rotation velocity of the solar neighborhood around the Galactic center $V_0,$ and the geometric characteristics of the spiral pattern or the kinematic parameters of the spiral density wave were estimated from various samples of masers.

By now almost all of the results of VLBI trigonometric parallax measurements for masers have been obtained with instruments located in the Earth's Northern Hemisphere. Therefore, although these sources are distributed over a huge Galactic space, there is still an empty zone in the fourth Galactic quadrant. However, the measurements are already
enough to successfully estimate, for example, the pitch angles of the spiral arms (Reid et al. 2019) or $R_0$ (Nikiforov and Veselova 2018) from the spatial distribution.

The radio astronomers performing VLBI observations of masers persistently note the presence of deviations from the circular motions of these objects with a value of 10--20~km s$^{-1}$ (Wu et al. 2019; Immer et al. 2019; Sakai et al. 2019; Hirota et al. 2020; Xu et al. 2021). Many masers are known to be associated with binary star systems, with expanding, often asymmetric, envelopes around stars. It may also well be that they belong to runaway stars. This all increases the observed velocity dispersion of these objects. However, we associate (Bobylev and Bajkova 2010; Bobylev et al. 2010) the presence of systematic deviations from the circular motions of the youngest stars in general and masers in particular with the influence of the Galactic spiral density wave.

The goal of this paper is to redetermine the Galactic rotation and Galactic spiral density wave parameters using up-to-date data on masers and radio stars
with measured VLBI trigonometric parallaxes and
proper motions. To achieve this goal, we use the most accurate data--masers with relative parallax errors less than 10\%.

 \subsection*{METHODS}
From observations for each star we have the line-of-sight velocity $V_r$ and the two tangential velocity components $V_l=4.74r\mu_l\cos b$ qnd $V_b=4.74r\mu_b$
along the Galactic longitude $l$ and latitude $b,$ respectively, 4.74 is the dimension coefficient. All three velocities are expressed in km s$^{-1}$ and $r$ is the stellar
heliocentric distance in kpc. The proper motion components $\mu_l\cos b$ and $\mu_b$ are expressed in mas yr$^{-1}$ (milliarcseconds per year). The velocities $U, V, W$ directed along the rectangular Galactic coordinate axes are calculated via the components $V_r, V_l, V_b:$
 \begin{equation} \begin{array}{lll}
 U=V_r\cos l\cos b-V_l\sin l-V_b\cos l\sin b,\\
 V=V_r\sin l\cos b+V_l\cos l-V_b\sin l\sin b,\\
 W=V_r\sin b                +V_b\cos b,
 \label{UVW} \end{array} \end{equation}
where the velocity $U$ is directed from the Sun toward the Galactic center or, more precisely, to the Galactic rotation axis, $V$ is in the direction of Galactic rotation,
and $W$ is directed to the north Galactic pole. We
can find two velocities, $V_R$ directed radially away from
the Galactic center and $V_{circ}$ orthogonal to it pointing
in the direction of Galactic rotation, based on the following relations:
 \begin{equation} \begin{array}{lll}
  V_{circ}= U\sin \theta+(V_0+V)\cos \theta, \\
       V_R=-U\cos \theta+(V_0+V)\sin \theta,
 \label{VRVT}
 \end{array} \end{equation}
where the position angle $\theta$  obeys the relation $\tan\theta=y/(R_0-x)$, $x, y, z$ are the rectangular stellar heliocentric coordinates (the velocities $U, V, W$ are directed
along the corresponding $x, y, z$ axes), and $V_0$ is the linear rotation velocity of the Galaxy at the solar distance $R_0.$

\subsubsection*{Determination of Galactic Rotation Parameters}
To determine the parameters of the Galactic rotation curve, we use the equations derived from Bottlinger's formulas, in which the angular velocity $\Omega$ is expanded into a series to terms of the second order of smallness in $r/R_0:$
\begin{equation} \begin{array}{lll}
 V_r=-U_\odot\cos b\cos l-V_\odot\cos b\sin l-W_\odot\sin b\\
 +R_0(R-R_0)\sin l\cos b\Omega^\prime_0
 +0.5R_0(R-R_0)^2\sin l\cos b\Omega^{\prime\prime}_0,
 \label{EQ-1}
 \end{array} \end{equation}
\begin{equation} \begin{array}{lll}
 V_l= U_\odot\sin l-V_\odot\cos l-r\Omega_0\cos b\\
 +(R-R_0)(R_0\cos l-r\cos b)\Omega^\prime_0
 +0.5(R-R_0)^2(R_0\cos l-r\cos b)\Omega^{\prime\prime}_0,
 \label{EQ-2}
 \end{array}\end{equation}
\begin{equation} \begin{array}{lll}
 V_b=U_\odot\cos l\sin b + V_\odot\sin l \sin b-W_\odot\cos b\\
 -R_0(R-R_0)\sin l\sin b\Omega^\prime_0
    -0.5R_0(R-R_0)^2\sin l\sin b\Omega^{\prime\prime}_0,
 \label{EQ-3}
 \end{array} \end{equation}
where $R$ is the distance from the star to the Galactic rotation axis,
$R^2=r^2\cos^2 b-2R_0 r\cos b\cos l+R^2_0.$ The velocities $(U,V,W)_\odot$ are the mean group velocity of the sample; they reflect the peculiar motion of the Sun and, therefore, are taken with opposite signs. $\Omega_0$ is the angular velocity of Galactic rotation at the solar distance $R_0,$ the parameters $\Omega^{\prime}_0$ and $\Omega^{\prime\prime}_0$ are the corresponding derivatives of the angular velocity, and $V_0 = R_0\Omega_0.$

We take $R_0$ to be $8.1\pm0.1$~kpc. This value was derived as a weighted mean of a large number of present-day individual estimates in Bobylev and Bajkova (2021). It is interesting to note the highly accurate present-day individual measurement of $R_0$ obtained by Abuter et al. (2019) by analyzing a 16-year series of observations of the motion of the star S2
around the massive black hole Sgr\,A* at the Galactic center, $R_0=8.178\pm0.013$\,(stat)$\pm0.022$\,(syst)~kpc. However, instrumental aberrations were shown to be present in the latter publication of this team (Abuter
et al. 2021). Therefore, all the previous estimates of the collaboration since 2018 were revised and the refined $R_0=8.275\pm0.009\pm0.033$~kpc was proposed.

A solution of the conditional equations (3)--(5) is sought by the least-squares method (LSM). As a result, we obtain an estimate of the following six unknowns: $(U,V,W)_\odot,$ $\Omega_0$, $\Omega^{\prime}_0,$ and $\Omega^{\prime\prime}_0$. Note that the velocities $U, V,$ and $W$ in Eqs. (2) were freed from the peculiar solar velocity $U_\odot,$ $V_\odot,$ and $W_\odot$ with the values found through the LSM solution of the kinematic equations (3)--(5).

\subsubsection*{Determination of Spiral DensityWave Parameters}
The influence of the spiral density wave in the radial, $V_R,$ and residual tangential, $\Delta V_{circ}$, velocities is periodic with an amplitude $\sim$10--15 km s$^{-1}.$
However, refining the specific values of the perturbation amplitudes is one of the objectives of this study. According to the linear density wave theory by Lin
and Shu (1964), the velocity perturbations satisfy the following relations:
 \begin{equation}
 \begin{array}{lll}
       V_R =-f_R \cos \chi,\\
 \Delta V_{circ}= f_\theta \sin\chi,
 \label{DelVRot}
 \end{array}
 \end{equation}
where
 \begin{equation}
 \chi=m[\cot(i)\ln(R/R_0)-\theta]+\chi_\odot
 \end{equation}
is the phase of the spiral density wave ($m$ is the number of spiral arms, $i$ is the pitch angle of the spiral pattern, $i<0$ for a wound spiral, $\chi_\odot$ is the radial
phase of the Sun in the spiral density wave; $f_R$ and $f_\theta$ are the amplitudes of the radial and tangential velocity perturbations, which are assumed to be positive.

We reveal periodicities in the velocities $V_R$ and $\Delta V_{circ}$ based on a spectral (periodogram) analysis, which is described in Bajkova and Bobylev (2012).
The wavelength $\lambda$ (the distance between adjacent spiral arm segmentsmeasured along the radial direction) is calculated based on the relation
\begin{equation}
 2\pi R_0/\lambda=m\cot(|i|).
 \label{a-04}
\end{equation}
Let there be a series ofmeasured velocities $V_{R_n}$ (these can be the radial, $V_R,$ tangential, $\Delta V_{circ}$, or vertical, $W,$ velocities), $n=1,\dots,N$, where $N$ is
the number of objects. The objective of the spectral analysis is to extract a periodicity from the data series in accordance with the adopted model describing a spiral density wave with parameters $f,$ $\lambda$~(or $i),$ and $\chi_\odot$.

Having taken into account the logarithmic behavior of the spiral density wave and the position angles of the objects $\theta_n$, our spectral analysis of the series
of velocity perturbations is reduced to calculating the
square of the amplitude (power spectrum) of the standard
Fourier transform (Bajkova and Bobylev 2012):
\begin{equation}
 \bar{V}_{\lambda_k} = \frac{1} {N}\sum_{n=1}^{N} V^{'}_n(R^{'}_n)
 \exp\biggl(-j\frac {2\pi R^{'}_n}{\lambda_k}\biggr),
 \label{29}
\end{equation}
where $\bar{V}_{\lambda_k}$ is the $k$th harmonic of the Fourier transform with wavelength $\lambda_k=D/k$, $D$ is the period of the series being analyzed,
 \begin{equation}
 \begin{array}{lll}
 R^{'}_{n}=R_0\ln(R_n/R_0),\\
 V^{'}_n(R^{'}_n)=V_n(R^{'}_n)\times\exp(jm\theta_n).
 \label{21}
 \end{array}
\end{equation}
The sought-for wavelength $\lambda$ corresponds to the peak value of the power spectrum Speak. The pitch angle of the spiral density wave is found from Eq.~(8). We determine the perturbation amplitude and phase by fitting the harmonic with the wavelength found to the observational data. To estimate the perturbation amplitude, we can also used the relation
 \begin{equation}
 f_R (f_\theta, f_W)=2\times\sqrt{S_{peak}}.
 \label{Speak}
 \end{equation}
In our spectral analysis we estimated the errors of the sought-for parameters through Monte Carlo simulations by performing 100 cycles of computations. For
this number of cycles the mean values of the solutions
virtually coincide with the solutions obtained from the
original data without adding any measurement errors.
Measurement errors were added to the velocities $V_R,$ $\Delta V_{circ}$, and $W$ and well as to the errors in the coordinates of the sources $x, y,$ and $z.$

 \subsection*{DATA}
The maser sources are stars with extended gas-dust envelopes in which the pumping effect arises. Both young stars and protostars of various masses and old stars, for example, Miras, possess the masing effect. In this paper we use the observations of only young stars closely associated with active starforming regions.

The main data sources in this paper are the catalogues by Reid et al. (2019) and Hirota et al. (2020). The list by Reid et al. (2019) includes data on
199 masers. The VLBI observations were carried out
at several radio frequencies within the BeSSeL (The
Bar and Spiral Structure Legacy Survey\footnote{http://bessel.vlbi-astrometry.org}) project.
Hirota et al. (2020) presented a catalogue of 99 maser
sources that were observed exclusively at 22 GHz
within the VERA (VLBI Exploration of Radio Astrometry\footnote{http://veraserver.mtk.nao.ac.jp})
program. The lists by Reid et al. (2019) and Hirota et al. (2020) have a high percentage of
common measurements and, therefore, we compiled
a list of data without coincidences. In addition, we
added several new parallax determinations for a number
of masers after 2020 (Sakai et al. 2020; Ortiz-Le\'on et al. 2020; Xu et al. 2021; Sakai et al. 2022; Bian et al. 2022). Apart from the masers, there are some number of radio stars in our list, which are very young stars located mostly in the Gould Belt
region. The observations of these stars we performed
by VLBI in continuum (Ortiz-Le\'on et al. 2018; Galli
et al. 2018). Currently, our list contains a total of 267 determinations of the VLBI parallaxes and proper motions for masers and radio stars.

In this paper we use sources with relative parallax errors less than 10\%. This approach is related to the fact that, for example, the Lutz-Kelker (1973) bias should be taken into account at relative errors of the measured parallaxes more than 10\%. To apply this
bias, the distribution of matter in the Galaxy needs to
be known well. In contrast, the actual distribution of
matter at great distances from the Sun, where most
of the masers are located, is currently not known well enough to solve this problem. The corrections for this bias are model-dependent and, therefore, the need for them arises only in extreme cases. As shown in Stepanishchev and Bobylev (2013), in isolated instances (at $\sigma_\pi/\pi\gg10\%$) the corrections for masers can be significant.

\begin{figure}[t]
{ \begin{center}
  \includegraphics[width=0.85\textwidth]{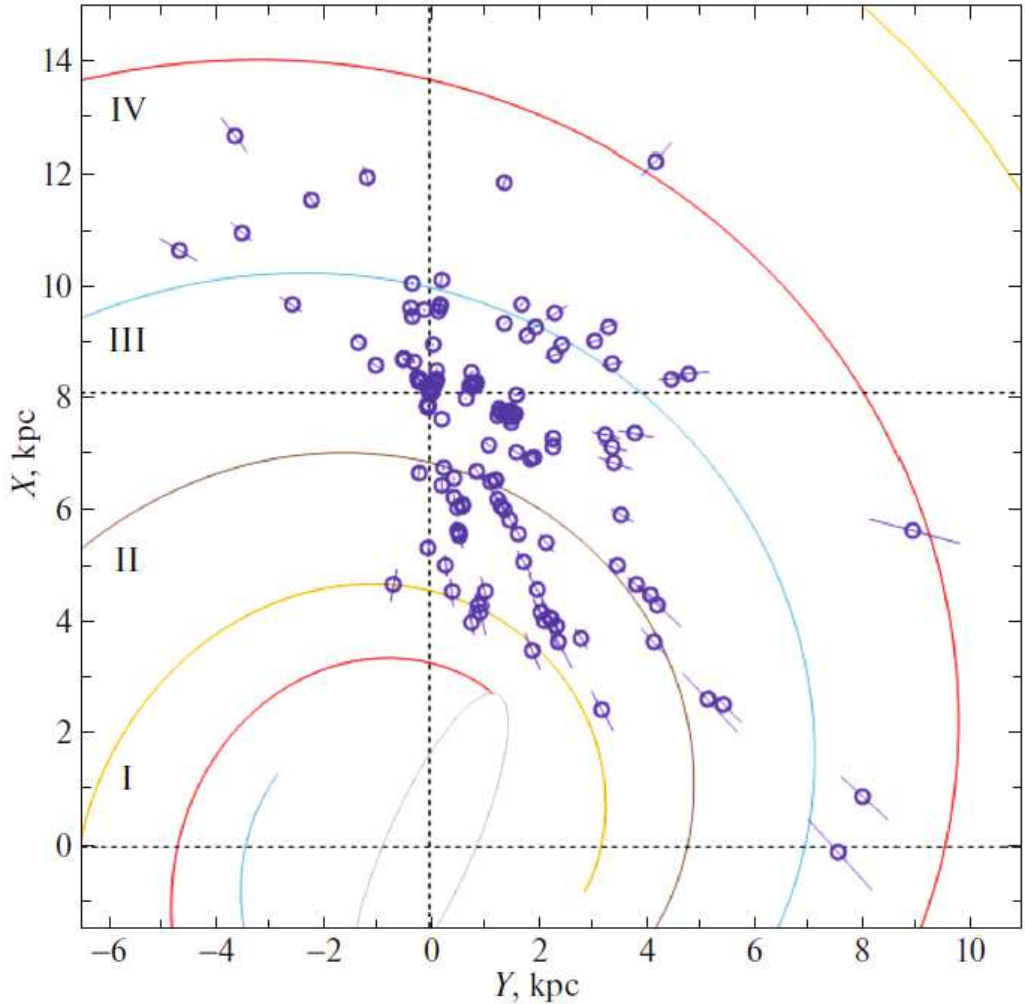}
  \caption{
Distribution of 150 masers and radio stars with trigonometric parallax errors less than 10\% in projection onto the Galactic $XY$ plane, the four-armed spiral pattern with the pitch angle $i=-13^\circ$ is shown, the central Galactic bar is marked. 
  }
 \label{f-150-XY}
\end{center}}
\end{figure}

The selection of sources was made under the condition $R>4$~kpc. This is because the central bar exerts a strong gravitational influence on the kinematics of stars in the region $R<4$~kpc. This gives rise to significant deviations from the circular orbits and
increases the observed stellar space velocity dispersion.
Therefore, when seeking the Galactic rotation
parameters, the region $R<4$~kpc is usually (Reid
et al. 2019) excluded from consideration. Moreover, the spiral pattern begins from the ends of the central bar and, hence, there is no need to use objects from the region $R<4$~kpc when seeking the spiral density wave parameters as well.

To get rid of the great outliers, we use the following constraints on the space velocities:
 \begin{equation}
 \begin{array}{rcl}
       |U|<90~\hbox{km s$^{-1}$},\\
       |V|<90~\hbox{km s$^{-1}$},\\
       |W|<60~\hbox{km s$^{-1}$},
  \label{cut}
 \end{array}
 \end{equation}
where $U, V,$ and $W$ are the residual velocities, since the peculiar solar velocity and the Galactic rotation velocity were presubtracted from them. We use the Galactic rotation curve whose parameters were found by analyzing masers in Bobylev et al. (2020) as the
preknown one. Five or six stars with the greatest
outliers are rejected as a result of applying the criteria. On the whole, however, we use the 3$\sigma$ rejection criterion when seeking the LSM solution of the system of conditional equations (3)--(5).

The distribution of the available masers with relative parallax errors less than 10\% selected under the condition $R>4$~kpc is presented in Fig.~1. The
number of such objects is 150. In this figure we
use the coordinate system in which the $X$ axis is
directed from the Galactic center to the Sun and the
direction of the $Y$ axis coincides with the direction of
Galactic rotation. The four-armed spiral pattern with the pitch angle $i=-13^\circ$, as inferred by Bobylev and Bajkova (2014), is presented; here it was constructed with $R_0=8.1$~kpc, the Roman numerals number the following four spiral arms: Scutum (I), Carina–Sagittarius (II), Perseus (III), and the Outer Arm (IV).

 \subsection*{RESULTS AND DISCUSSION}
 \subsubsection*{Galactic Rotation Parameters}
Three approaches were used in seeking the LSM solution of the system of conditional equations (3)--(5). In the first case, all equations were used, i.e.,
all three velocities were involved: $V_r$, $V_l,$ and $V_b$. In
the second case, the two equations (4) and (5), but
involving the maser proper motions, were used. In the third case, only one equation (4) involving only one velocity component $V_l$ was used.

 \begin{table}[t] \caption[]{\small
The kinematic parameters found from 150 masers located in the Galactic region $R>4$~kpc }
  \begin{center}  \label{t:150}
  \small\begin{tabular}{|l|r|r|r|r|r|}\hline
        Parameters  & $V_r+V_l+V_b$  &      $V_l+V_b$ &          $V_l$ \\\hline
   $U_\odot,$ km s$^{-1}$ & $ 9.15\pm0.86$ & $ 9.23\pm1.46$ & $ 8.89\pm1.56$ \\
   $V_\odot,$ km s$^{-1}$ & $12.81\pm0.86$ & $11.15\pm1.01$ & $11.22\pm1.04$ \\
   $W_\odot,$ km s$^{-1}$ & $ 8.93\pm0.75$ & $ 8.47\pm0.67$ &            --- \\
      $\Omega_0,$ km s$^{-1}$ kpc$^{-1}$ & $ 30.18\pm0.38$ & $ 29.46\pm0.43$ & $ 29.39\pm0.45$  \\
  $\Omega^{'}_0,$ km s$^{-1}$ kpc$^{-2}$ & $-4.368\pm0.077$& $-3.914\pm0.108$& $-3.906\pm0.110$ \\
 $\Omega^{''}_0,$ km s$^{-1}$ kpc$^{-3}$ & $ 0.845\pm0.037$& $ 0.652\pm0.045$& $ 0.653\pm0.046$ \\
      $\sigma_0,$ km s$^{-1}$ &          8.9    &             8.1 &              8.5 \\
           $V_0,$ km s$^{-1}$ &   $244.4\pm4.3$ &   $238.6\pm4.6$ &$238.1\pm4.7$ \\\hline
\end{tabular}\end{center} \end{table}

The three approaches to using the conditional equations (3)--(5) are applied for the following reasons. First of all, it should be noted that VLBI observations
give the original trigonometric parallaxes and proper motions of the masers or radio stars. The systematic line-of-sight velocities of the objects are
taken from other sources. Therefore, it is interesting
to compare the results obtained with and without
the involvement of the line-of-sight velocities. A
peculiarity in using the components $V_l$ and $V_b$ is
that the velocities $V_l$ make the dominant contribution
to the estimate of almost all sought-for kinematic
parameters. However, the velocity $W_\odot$ cannot be
determined when using only the component $V_l.$ As can be seen from Eq.~(5), the coefficients before the sought-for parameters $(U,V)_\odot,$ $\Omega_0$, $\Omega^{\prime}_0,$ and $\Omega^{\prime\prime}_0$
contain the term $\sin b$ very close to zero for distant
masers and $\cos b$ appears only at the unknown $W_\odot$.
Therefore, it is interesting to compare the solutions found with and without the involvement of the velocities $V_b.$ Note that our sample contains quite a few nearby sources with relatively large angles $b.$

The kinematic parameters found by all three approaches from 150 masers located in the Galactic region $R>4$~kpc are given in Table~1. Figure~2 presents the radial velocities $V_R,$ rotation velocities $V_{circ}$, and residual rotation velocities $\Delta V_{circ}$ for our sample of 150 masers.

It can be seen from Fig. 2 that the radial and tangential velocities of a small group of stars in the Gould Belt region ($r<0.5$~kpc) ``stick out''. To
estimate the influence of this group of stars on the
parameters of the general solution, we obtained the joint ($V_r+V_l+V_b$) solution by excluding the Gould Belt region ($r<0.5$~kpc). With this approach, using 119 masers, we found the velocities $(U,V,W)_\odot=(5.02,13.86,8.77)\pm(1.21,1.22,0.93)$~km s$^{-1}$ and
\begin{equation}  \label{sol-150} \begin{array}{lll}
      \Omega_0 =~29.73\pm0.48 ~\hbox{km s$^{-1}$ kpc$^{-1}$},\\
  \Omega^{'}_0 =-4.246\pm0.089~\hbox{km s$^{-1}$ kpc$^{-2}$},\\
 \Omega^{''}_0 =~0.826\pm0.043~\hbox{km s$^{-1}$ kpc$^{-3}$},
 \end{array} \end{equation}
where the error per unit weight is $\sigma_0=9.9$~km s$^{-1}$
and $V_0=240.8\pm4.9$~km s$^{-1}$ (for the adopted $R_0=8.1\pm0.1$~kpc).

\begin{figure}[t]
{ \begin{center}
  \includegraphics[width=0.75\textwidth]{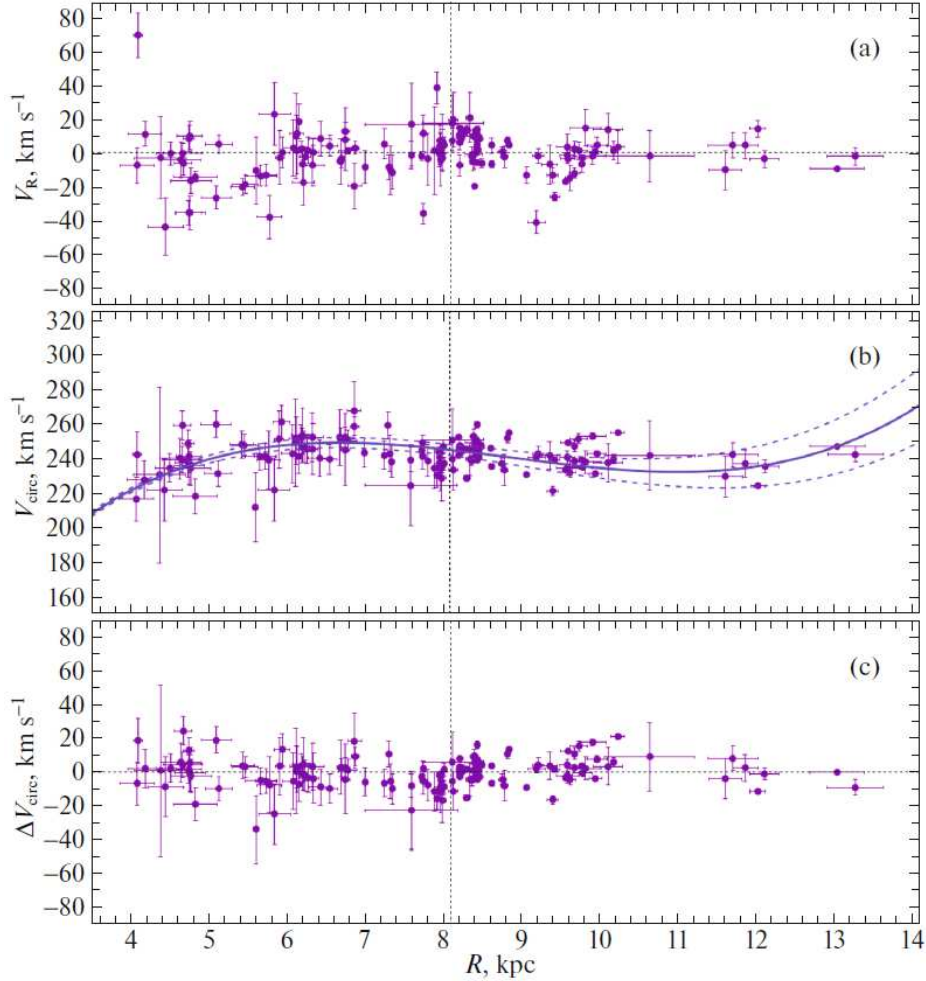}
  \caption{
Radial velocities $V_R$~(a), rotation velocities $V_{circ}$~(b), and residual rotation velocities $\Delta V_{circ}$~(c) of the masers versus distance $R.$ The rotation curve found from these sources with an indication of the boundaries of the 1$\sigma$ confidence regions is presented; the vertical line marks the Sun's position.}
 \label{f-150-rotation}
\end{center}}
\end{figure}
 \begin{table}[t] \caption[]{\small
The kinematic parameters found from 104 masers located no farther than 3 kpc from the Sun }
  \begin{center}  \label{t:104}
  \small\begin{tabular}{|l|r|r|r|r|r|}\hline
        Parameters  & $V_r+V_l+V_b$  &      $V_l+V_b$ \\\hline
   $U_\odot,$ km s$^{-1}$ & $11.90\pm0.91$ & $12.21\pm1.85$ \\
   $V_\odot,$ km s$^{-1}$ & $14.14\pm0.93$ & $12.49\pm1.08$ \\
   $W_\odot,$ km s$^{-1}$ & $ 9.29\pm0.75$ & $ 8.79\pm0.72$ \\
      $\Omega_0,$ km s$^{-1}$ kpc$^{-1}$ & $ 33.5\pm1.0$ & $ 32.2\pm1.4 $\\
  $\Omega^{'}_0,$ km s$^{-1}$ kpc$^{-2}$ & $-4.55\pm0.16$& $-4.20\pm0.24$\\
 $\Omega^{''}_0,$ km s$^{-1}$ kpc$^{-3}$ & $ 1.20\pm0.13$& $ 0.99\pm0.14$\\
      $\sigma_0,$ km s$^{-1}$    &    $ 7.6$    &         $ 6.9$ \\
           $V_0,$ km s$^{-1}$    &    $271\pm9$ &     $261\pm12$ \\\hline
\end{tabular}\end{center} \end{table}
\begin{figure}[t]
{ \begin{center}
  \includegraphics[width=0.825\textwidth]{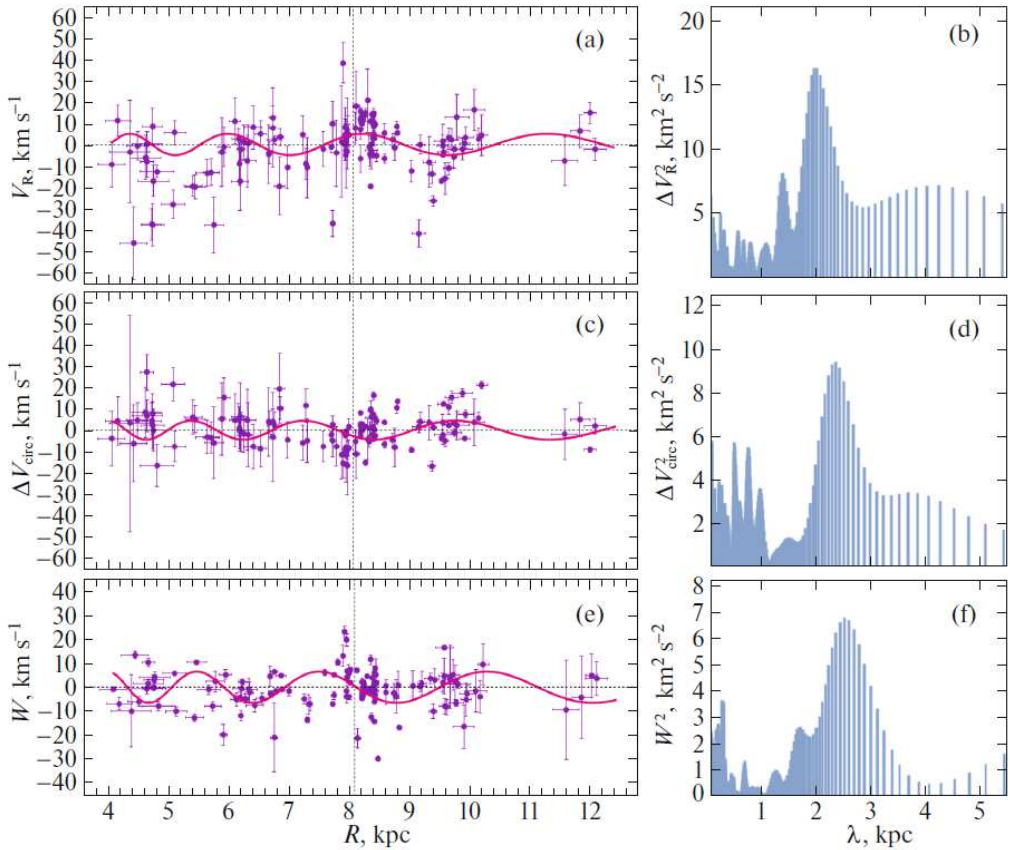}
  \caption{
Radial velocities $V_R$ versus distance $R$~(a) and their power spectrum (b), residual rotation velocities $\Delta V_{circ}$ versus distance $R$~(c) and their power spectrum (d), vertical velocities $W$ versus distance $R$~(e) and their power spectrum (f); the solid
wavy lines represent the results of our spectral analysis. Here, we used 134 masers with relative trigonometric parallax errors less than 10\% located no farther than 5~kpc from the Sun.
 }
 \label{f-134-ru}
\end{center}}
\end{figure}
\begin{figure}[t]
{ \begin{center}
  \includegraphics[width=0.95\textwidth]{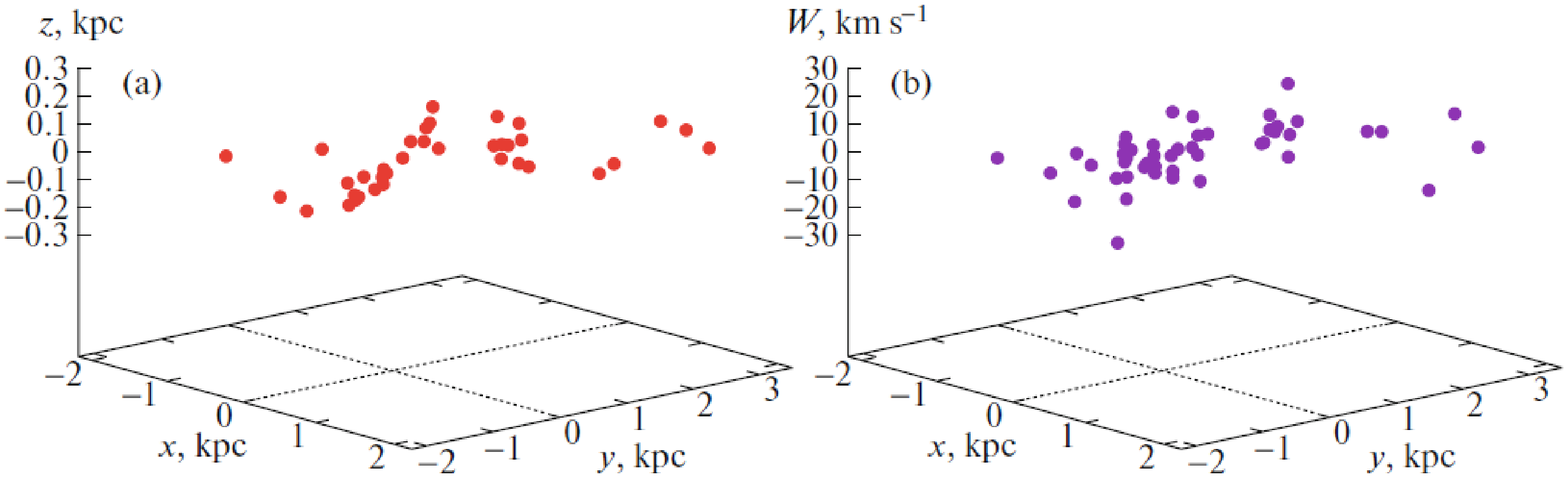}
  \caption{
Three-dimensional distribution of masers and radio stars belonging to the Local Arm (a) and their vertical velocities (b).
 }
 \label{f-radclif}
\end{center}}
\end{figure}

Rastorguev et al. (2017) considered a sample of 130 masers with measured trigonometric parallaxes. For example, based on the model of purely circular
disk rotation (Model A2), they found the group velocity
components $(U,V,W)_\odot=(7.73,17.69,8.64)\pm(1.52,1.20,0.91)$~km s$^{-1}$ and the following parameters of the Galactic rotation curve:
 $\Omega_0=29.03\pm0.52$~km s$^{-1}$ kpc$^{-1}$,
 $\Omega^{1}_0=-3.94\pm0.08$~km s$^{-1}$ kpc$^{-2}$,
 $\Omega^{2}_0= 1.13\pm0.07$~km s$^{-1}$ kpc$^{-3}$,
 $\Omega^{3}_0= 0.06\pm0.11$~km s$^{-1}$ kpc$^{-4}$,
 $\Omega^{4}_0= 0.30\pm0.02$~km s$^{-1}$ kpc$^{-5}$, and
 $\Omega^{5}_0=-0.14\pm0.02$~km s$^{-1}$ kpc$^{-6}$, where $V_0=241\pm10$~km s$^{-1}$ for the value of $R_0=8.31\pm0.13$~kpc found. It can be noted that the estimates of the
parameters that are given in Table~1 and the solution (13) were obtained with smaller errors. We achieved this by using a larger number of sources and by applying the constraint on the relative maser parallax error (less than 10\%) and the criteria (12).

Reid et al. (2019) performed a kinematic analysis of a sample of 147 masers. In contrast to our approach, these authors applied an expansion of the linear rotation (circular) velocity of the Galaxy into a series. As a result, they obtained estimates of $R_0=8.15\pm0.15$~kpc and $\Omega_\odot=30.32\pm0.27$~km s$^{-1}$ kpc$^{-1}$, where $\Omega_\odot$ is the angular velocity of precisely the Sun.

It is also interesting to note the results of our analysis of the proper motions and trigonometric parallaxes for a sample of 9750 OB2 stars that
were obtained in Bobylev and Bajkova (2022). One conditional equation (4) involving only one velocity component $V_l$ was used in seeking the LSM solution. As a result, we found the velocities $(U,V)_\odot=(7.17,7.37)\pm(0.16,0.24)$~km s$^{-1}$ and
     $\Omega_0 =29.700\pm0.076$~km s$^{-1}$ kpc$^{-1},$
  $\Omega^{'}_0=-4.008\pm0.022$~km s$^{-1}$ kpc$^{-2},$ and
 $\Omega^{''}_0=~0.671\pm0.011$~km s$^{-1}$ kpc$^{-3},$ where
 $V_0=240.6\pm3.0$~km s$^{-1}$ (for the adopted $R_0=8.1\pm0.1$~kpc).

There is good agreement of the listed estimates of the Galactic rotation parameters obtained by various authors with the results presented in Table 1 and
the solution (13). In contrast, there is no complete
agreement with regard to the velocities $U_\odot$ and $V_\odot$.

The error per unit weight $\sigma_0$ that we find by seeking the LSM solution of the kinematic equations is calculated as a weighted mean of the residuals. Thus,
this is a quantity averaged over three stellar velocity dispersion directions. It can be seen from Table 1 and the solution (13) that both when using only the proper
motions of the masers and when using their line-of-sight
velocities, the values of $\sigma_0$ are comparable and
do not exceed 10~km s$^{-1}$. Here, the constraints (12) perform their role, but the constraints on the relative parallax errors mainly operate.

For example, Bobylev et al. (2020) analyzed a sample of 239 masers (almost all of the then available measurements) without using any special constraints, where $\sigma_0$ was 12.8~km s$^{-1}$, which is more likely typical for older stars, for example, classical Cepheids.

In this case, we obtained the following estimates of the group velocity for the sample of masers:
 $U_\odot= 7.79_{-1.27}^{+1.23}$~km s$^{-1}$,
 $V_\odot=15.04_{-1.25}^{+1.24}$~km s$^{-1}$,
 $W_\odot= 8.57_{-1.23}^{+1.18}$~km s$^{-1}$, and the parameters of the Galactic
rotation curve:
          $\Omega_0=29.01_{-0.34}^{+0.33}$~km s$^{-1}$ kpc$^{-1},$
   $\Omega^{'}_0=-3.901_{-0.069}^{+0.068}$~km s$^{-1}$ kpc$^{-2},$
  $\Omega^{''}_0= 0.831_{-0.032}^{+0.032}$~km s$^{-1}$ kpc$^{-3},$
 $\Omega^{'''}_0=-0.104_{-0.019}^{+0.018}$~km s$^{-1}$ kpc$^{-4},$ we
also found $R_0=8.15_{-0.20}^{+0.04}$~kpc.  Here, we can also note that in this paper the group velocity components and the Galactic rotation parameters are estimated
with smaller errors.

Next, we produced a local sample of 104 masers selected under the constraint on the heliocentric distance $r<3$~kpc. The kinematic parameters found
from this sample are given in Table 2. The conditional
equations were solved in two ways. Here, the velocities
$V_b$ were not excluded, since the stars are nearby
ones and, therefore, have heights b sufficient for a
reliable determination of the sought-for unknowns of
the model. We can see an increased value of $\Omega_0$
compared the ones found above from the samples
of more distant masers. This effect is most likely
caused by the influence of the masers and radio stars belonging to the Gould Belt structure. The main result obtained with this approach is that the error per unit weight $\sigma_0=6.9$~km s$^{-1}$ found in the last column of Table 2 is very small. This suggests that we are dealing with a population of very young objects.
When using increasingly distant samples of stars, the value of this quantity increases, which is explained by the joint influence of the stellar trigonometric parallax
and proper motion errors.

\begin{figure}[t]
{ \begin{center}
  \includegraphics[width=0.825\textwidth]{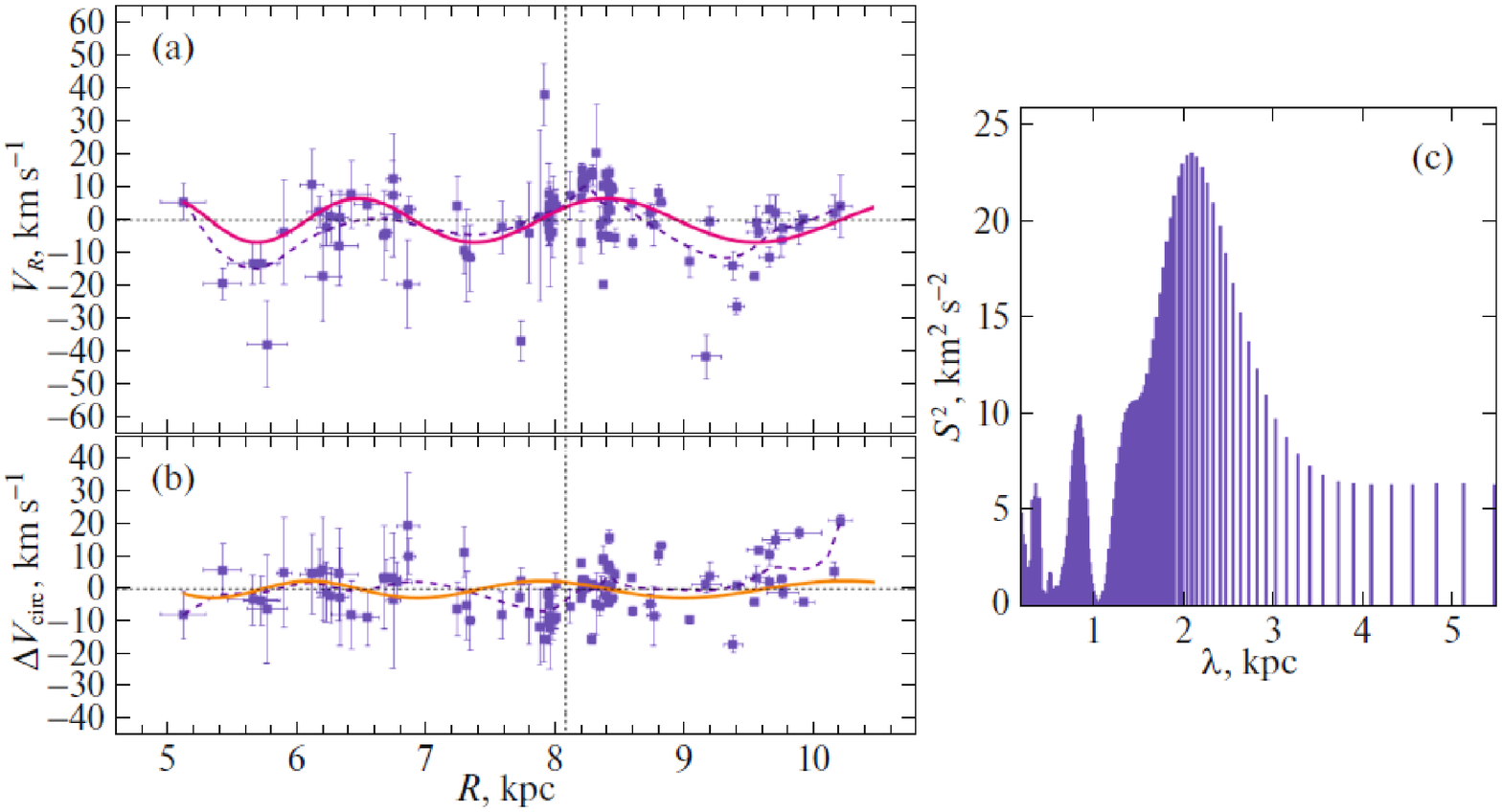}
  \caption{
Radial velocities $V_R$~(a) and residual rotation velocities $\Delta V_{circ}$~(b) versus distance $R$ and their power spectrum, the data averaging is indicated by the dashed lines; the solid wavy lines represent the results of our spectral analysis. Here, we used 104 masers with relative trigonometric parallax errors less than 10\% located no farther than 3~kpc from the Sun.
 }
 \label{f-104-ru}
\end{center}}
\end{figure}

 \subsubsection*{Spectral Analysis}
{\bf Separate solutions.} For our spectral analysis we
used data on 134 masers within 5~kpc of the Sun.
Here, we slightly reduced the sample size for the
following reasons. According to the linear density
wave theory by Lin and Shu (1964), the amplitudes
of the radial and tangential velocity perturbations are
not constant. Indeed, the spiral pattern rotates rigidly.
Therefore, in the corotation region the difference between
the linear rotation velocities of the Galaxy and
the spiral pattern is zero, while near the ends of the
central bar it is maximal. The effect of an enhanced
velocity dispersion at $R\sim4$~kpc can be noticed in
Figs. 2a and 2b, which is particularly pronounced in
the radial velocities. When analyzing the tangential
velocities, it is useful to reduce the sample radius---it is better to use the $R$ interval where the Galactic rotation curve is closest to the flat one.

In our separate spectral analysis of the radial,
$V_R,$ residual tangential, $\Delta V_{circ}$, and vertical maser
velocities we obtained the following estimates:
      $f_R=8.1\pm1.4$~km s$^{-1}$,
 $f_\theta=6.1\pm1.7$~km s$^{-1}$,
      $f_W=5.2\pm1.5$~km s$^{-1}$,
 $\lambda_R=2.1\pm0.3$~kpc, $\lambda_\theta=2.4\pm0.8$~kpc, and $\lambda_W=2.6\pm0.7$~kpc,
 $(\chi_\odot)_R=-158\pm15^\circ$, $(\chi_\odot)_\theta=-140\pm18^\circ,$ and  $(\chi_\odot)_W=-180\pm18^\circ$.
The results of this analysis are presented in Fig. 3.

It is interesting to note the significance (sig) of
the main peak in each of the cases marked in Fig. 3:
sig$_R=0.9997$, sig$_\theta=0.9408,$ and sig$_W=0.8386$.
These values suggest that the parameters of the spiral
density wave were determined most reliably from the
radial velocities $V_R$ and least reliably from the vertical
velocities $W.$

At the same time, the presence of a periodicity in
the vertical maser velocities is an important result of
this paper. The point is that in the classical density
wave theory by Lin and Shu (1964) the vertical stellar
velocities are not considered. Based on a spectral
analysis, Bobylev and Bajkova (2015) for the first time
found a periodicity in the vertical maser velocities with
the wavelength $\lambda_W=3.4\pm0.7$~kpc and the amplitude
$f_W=4.3\pm1.2$~km s$^{-1}$.

Currently, vertical velocity perturbations of various nature are known to be present in the Galactic disk (L\'opez-Corredoira et al. 2014; Widrow et al. 2014; Antoja et al. 2018; Wang et al. 2020; Thulasidharan et al. 2021). These can be the large-scale perturbations associated with the Galactic Warp caused both
by the gravitational influence of some external factor
and by a nongravitational one, for example, the fall to
the disk or a close flyby of a dwarf companion galaxy
of the Milky Way or a massive clump of dark matter.
There can also be the vertical velocity perturbations
associated with the natural disk oscillations or with
the influence of large-scale magnetic fields, etc.

Note that the Radcliffe wave (Alves et al. 2020) propagating along the Local Arm has recently been detected in the local region while analyzing the distribution
of molecular clouds. In the opinion of Alves et al. (2020), the wave has a wavelength of about 2~kpc with an amplitude of about 160~pc and is a damped one. The origin of this wave has not yet been established.

We selected masers and radio stars belonging to the Local Arm. These sources are excellently seen near the Sun in Fig.~1 as a fairly narrow chain. During the selection we assumed that the sources of the Local Arm were no farther than 3.5~kpc from the Sun and were bounded by two parallel lines: $x=0.286y-0.8$ and $x=0.286y+0.3.$ There were a total of 68 sources in this sample, the overwhelming majority of which are low-mass T\,Tauri stars. Figure~4 was constructed from the data on the selected stars. The Radcliffe wave is excellently seen in the stellar positions in Fig.~4a. No pronounced wave is observed in the vertical velocity distribution of the selected stars (Fig.~4b).

As a result, we are inclined to think that the wave
in the vertical maser velocities is more likely associated
with a larger-scale phenomenon, for example,
the influence of the spiral density wave (Fig. 3e). The
paper by Martinez-Medina et al. (2022) can also be
noted, where large-scale features other than noise
were shown to be present in the vertical velocity distribution of young stars from the Gaia EDR3 catalogue associated with the Galactic spiral structure. The most pronounced feature was noted by these authors at $R\sim9.5$~kpc, i.e., elongation in the Perseus arm region.

{\bf Joint solution.} In our joint spectral analysis of the radial, $V_R,$ and residual tangential, $\Delta V_{circ}$, velocities we assume that there is one value of the wavelength and the Sun's phase in the density wave for both types of velocities.

Using data on 134 masers within 5~kpc of the Sun with relative trigonometric parallax errors less than 10\%, we obtained the following estimates: $f_R=6.1\pm1.3$~km s$^{-1}$, $f_\theta=3.2\pm1.4$~km s$^{-1}$, and $\lambda=2.1\pm0.3$~kpc, $\chi_\odot=-141\pm15^\circ$. The significance of the main peak in the power spectrum was found to be sig=0.99991.

Note that the spectral analysis being applied takes
into account both the logarithmic behavior of the
spiral density wave and the distribution of object positions
angles. In a logarithmic wave the wavelength is
a variable quantity---it increases with $R.$ In this case,
the estimate of the parameters $\lambda, f_{\{R,\theta,W\}}, \chi_\odot$ refers
to the local region—the region near the Sun. As can
be seen from the distribution of masers in projection
onto the Galactic $XY$ plane in Fig.~1, two spiral
arm segments, Carina–Sagittarius and Perseus, are
essentially represented in the sample. Therefore, we
decided to perform a spectral analysis based also on
the local sample that encompasses this region. To
form the residual rotation velocities, we used the
Galactic rotation curve whose parameters were found
from our sample of 150 masers (Fig. 2b). As a result,
we obtained the following estimates from 104 masers:
  \begin{equation}
 \label{sol-104}
 \begin{array}{lll}
        f_R= 6.7\pm1.1~\hbox{km s$^{-1}$},\\
   f_\theta= 2.6\pm1.2~\hbox{km s$^{-1}$},\\
    \lambda= 2.1\pm0.3~\hbox{kpc},\\
 \chi_\odot=-148\pm15^\circ.
 \end{array}
 \end{equation}
The results of this spectral analysis are presented in
Fig.~5, where the significance of the main peak in the
power spectrum is sig=0.99999. Here, the value of
sig is slightly higher than that in the previous case
found by analyzing a wider solar neighborhood.

 \begin{table}[t] \caption[]{\small
The parameters of the Galactic spiral density wave estimated by various authors
 }
  \begin{center}  \label{t:03}
  \small
  \begin{tabular}{|l|c|c|c|c|c|c|c|}\hline
 Sample & Ref & $f_R,$ km s$^{-1}$ & $f_\theta,$ km s$^{-1}$ & $\lambda,$ kpc & $|i|,$ deg.
   & $\chi_\odot,$ deg. & $m$ \\\hline

 OB st., Ceph., OSCs & [1]& $3.6\pm0.4$ &$4.7\pm0.6$ & $$   &$ 4.2\pm0.2$ &$-165\pm1$& 2\\
 Cepheids      & [2]& $6.3\pm2.4$ &$4.4\pm2.4$ &   $$      &$ 6.8\pm0.7$  &$-70\pm16$& 2\\
 Cepheids      & [3]& $3.5\pm1.7$ &$7.5\pm1.8$ &   $$      &$11.4     $  &$-20\pm9$ & 4\\
 OB associations& [4]& $6.6\pm1.4$ &$1.8\pm1.4$ &$2.0\pm0.2$&  $$          &$$        &  \\
 Cepheids      & [5]& $6.7\pm2.3$ &$1.4\pm1.6$ &   $$      &$ 6.0\pm0.7$ &$-85\pm15$& 2\\
 OSCs         & [5]& $5.5\pm2.3$ &$0.2\pm1.6$ &   $$      &$12.2\pm0.7$ &$-88\pm15$& 4\\
 OB stars     & [5]& $6.6\pm2.5$ &$0.4\pm2.3$ &   $$      &$ 6.6\pm0.9$ &$-97\pm18$& 2\\
 OSCs, HI, HII & [6]& $5.9\pm1.1$ &$4.6\pm0.5$ &$2.1\pm0.5$&  $$ & $-119~~~~~~$ & \\
 Masers       & [7]& $7.7\pm1.6$ &  $$ &   $2.2\pm0.3$   &$ 5.0\pm0.5$  &$-147\pm10$& 2\\
 Masers       & [8]& $6.9\pm1.4$ &$2.8\pm1.0$ &          &$10.4\pm0.3$ &$-125\pm10$& 4\\
 OB stars     & [9]& $7.1\pm0.3$ &$6.5\pm0.4$ &$2.8\pm0.2$ &$$ &$-128\pm6$&  4\\
 OSCs         &[10]& $4.6\pm0.7$ &$1.1\pm0.4$ &    $$      & $$ &  $$  & 4\\
 OB stars     &[11]& $4.8\pm0.7$ &$4.1\pm0.9$ &$2.1\pm0.2$ &$9.4\pm0.9$ &$-116\pm12$& 4\\
 This paper  &&$6.7\pm1.1$ &$2.6\pm1.2$ &$2.1\pm0.3$ &$9.4\pm1.0$ &$-148\pm15$& 4\\
 \hline
 \end{tabular}\end{center}
  {\small
 [1] Byl and Ovenden (1978); [2] Mishurov et al. (1997); [3]~Mishurov and Zenina (1999); [4] Mel'nik et al. (2001); [5] Zabolotskikh et al. (2002); [6]~Bobylev et al. (2008); [7] Bajkova and Bobylev (2012); [8] Rastorguev et al. (2017); [9]~Bobylev and Bajkova (2018);
[10] Loktin and Popova (2019); [11]~Bobylev and Bajkova (2022).
  }
\end{table}

Apart from the periodicities found (thick lines), the
dashed lines in Fig. 5 indicate the averaged values of
the observer maser velocities. At great distance from
the Sun it makes no sense to compare these lines due
to the large position angles of some sources. However,
in the local solar neighborhood (encompassing
the two grand-design spiral arm segments closest
to the Sun) this comparison makes sense. As can
be seen from Fig. 5a, there is excellent agreement between the thick and dashed lines, suggesting that the values of $\lambda, f_R, \chi_\odot.$ found are reliable. In the tangential velocities we have much poorer agreement between these lines, but the amplitude $f_\theta$ here is also very small.

Table 3 presents the parameters of the spiral density
wave estimated from the most important tracers
of the spiral structure. As follows from this table, $f_\theta$
is usually smaller than $f_R.$

First of all, the results obtained with Cepheids
should be noted (Mishurov et al. 1997; Mishurov
and Zenina 1999; Zabolotskikh et al. 2002). They
are interesting in that the distances to these stars
define an independent distance scale, since they are
estimated based on the period–luminosity relation.
The random errors in the distances determined by this
method are 10--15\% (Berdnikov et al. 2000). The
most recent estimates of the distances to classical
Cepheids using photometric data and calibrations in
the near infrared suggest that the random errors are
about 5\% (Skowron et al. 2019).

The results obtained with OB stars (Byl and
Ovenden 1978; Zabolotskikh et al. 2002; Bobylev
and Bajkova 2018) are also of indubitable interest.
An important factor in estimating the kinematic
parameters is both the accuracy of the distances
and the accuracy of the proper motions. Some of
the present-day estimates given in the table were
obtained with data from the Gaia EDR3 catalogue
(Brown et al. 2021).

The distances to OB associations and open star
clusters (OSCs) are more accurate than those to
single stars. Therefore, the parameters found from
them must also be more reliable (Mel’nik et al. 2001;
Zabolotskikh et al. 2002; Bobylev et al. 2008; Loktin
and Popova 2019).

Finally, the table also contains the estimates
obtained using various samples of masers with measured
VLBI trigonometric parallaxes and proper
motions (Bajkova and Bobylev 2012; Rastorguev
et al. 2017).

 \subsection*{CONCLUSIONS}
We performed a kinematic analysis of a large sample
of Galactic masers and radio stars with measured
VLBI trigonometric parallaxes and proper motions.
We considered only masers with relative trigonometric
parallax errors less than 10\%.

When estimating the Galactic rotation parameters,
we considered various methods of solving the
basic kinematic equations—with and without the inclusion
of the line-of-sight velocities as well as with
and without the use of masers from the Gould Belt
region. Here, we were oriented to the error per unit
weight $\sigma_0,$ which may be considered as the maser
space velocity dispersion averaged over three directions.
While analyzing the sample of distant masers,
we showed that both when using only their proper
motions and when adding the line-of-sight velocities, the values of $\sigma_0$ are close between themselves, their values do to exceed 10~km s$^{-1}$.

As a result of the joint solution of the system
of kinematic equations using 150 masers from the
Galactic region $R>4$~kpc, we found the group velocity
components $(U,V,W)_\odot=(9.15,12.81,8.93)\pm(0.86,0.86,0.75)$~km s$^{-1}$ and the following parameters of the angular velocity of Galactic rotation:
 $\Omega_0=30.18\pm0.38$~km s$^{-1}$ kpc$^{-1},$
 $\Omega^{'}_0=-4.368\pm0.077$~km s$^{-1}$ kpc$^{-2},$ and
 $\Omega^{''}_0=0.845\pm0.037$~km s$^{-1}$ kpc$^{-3},$
where the error per unit weight $\sigma_0$ is 8.9~km s$^{-1}$ and $V_0=244.4\pm4.3$~km s$^{-1}$ for the adopted $R_0 = 8.1\pm0.1$~kpc. With the parameters of this Galactic rotation curve we find the residual tangential velocities.

We considered the kinematics of a local sample of
104 masers within 3~kpc of the Sun. The Galactic
rotation parameters were shown to be determined
from this sample with large errors. Here, there is a
strong influence of objects from the Gould Belt. In our
opinion, this leads to an overestimation of the angular
velocity of rotation, $\Omega_0=33.5\pm1.0$~km s$^{-1}$ kpc$^{-1},$
and, consequently, to an overestimated linear rotation
velocity, $V_0 = 271\pm9$~km s$^{-1}$. A positive effect of the
kinematic analysis of this sample of masers is a small
error per unit weight, $\sigma_0=6.9$~km s$^{-1}$. This means
that the masers and radio stars being considered here
are indeed representatives of the youngest population
of stars and protostars belonging to the Galactic thin
disk. For more distant masers the formal estimate of
their velocity dispersion increases due to the influence
of the measurement errors in their trigonometric parallaxes
and and the errors in their proper motions (the
random errors of the line-of-sight stellar velocities are
virtually independent of the distance).

To estimate the spiral density wave parameters,
we applied a spectral analysis of the velocities of
various samples of masers. For this purpose, we
used both separate and joint solutions. The separate
spectral analysis was applied to the radial, $V_R,$
residual tangential, $\Delta V_{circ}$, and vertical, $W,$ maser
velocities. The following estimates were obtained
from the data on 134 masers within 5~kpc of the Sun:
 $f_R=8.1\pm1.4$~km s$^{-1}$,
 $f_\theta=6.1\pm1.7$~km s$^{-1}$,
 $f_W=5.2\pm1.5$~km s$^{-1}$,
 $\lambda_R=2.1\pm0.3$~kpc, $\lambda_\theta=2.4\pm0.8$~kpc, and $\lambda_W=2.6\pm0.7$~kpc,
 $(\chi_\odot)_R=-158\pm15^\circ$,
 $(\chi_\odot)_\theta=-140\pm18^\circ,$ and
 $(\chi_\odot)_W=-180\pm18^\circ$.

We consider a logarithmic wave, when its wavelength increases with Galactocentric distance $R.$ Therefore, the $\lambda$ estimates refer to the solar neighborhood, where $\lambda$ is the distance between the Carina–Sagittarius and Perseus spiral arm segments.

It is important to note that the presence of periodic
perturbations first detected in Bobylev and Bajkova
(2015) was confirmed in the vertical maser velocities.
The perturbation amplitude $f_W$ and wavelength
$\lambda_W$ found are close to those derived from the
radial and residual tangential velocities. This argues
for the influence of the spiral density wave on the
vertical stellar velocities as well.

According to the classical linear theory of the Galactic spiral density wave by Lin and Shu (1964), one might expect one value of the wavelength $\lambda$ (consequently,
the Sun's phase in the wave $\chi_\odot$) in both
radial and residual tangential velocities. The vertical
velocities were not considered in this theory. In contrast,
here we see only satisfactory agreement in the
estimate of $\lambda_{R,\theta,W}$.

Interestingly, the local parameters of the spiral
density wave are determined quite well from the local
sample of 104 masers. For example, as a result of our
joint spectral analysis of the radial, $V_R,$ and residual
tangential, $\Delta V_{circ}$, maser velocities, we obtained
the following estimates:
 $f_R=6.7\pm1.1$~km s$^{-1}$,
 $f_\theta=2.6\pm1.2$~km s$^{-1}$, and
 $\lambda=2.1\pm0.3$~kpc, $\chi_\odot=-148\pm15^\circ$.

We considered a sample of 68 masers and radio stars belonging to the Local Arm. The presence of the Radcliffe wave in the spatial distribution of these sources was confirmed.

 \subsection*{ACKNOWLEDGMENTS}
We are grateful to Yu. N. Mishurov for the useful discussion of our results.

\bigskip \bigskip\medskip{\bf REFERENCES}{\small

1. R. Abuter, A. Amorim, N. Baub\"ock, et al. (GRAVITY
Collab.), Astron. Astrophys. 625, L10 (2019).

2. R. Abuter, A. Amorim,M. Baub\"ock, et al. (GRAVITY
Collab.), Astron. Astrophys. 647, A59 (2021).

3. J. Alves, C. Zucker, A. A. Goodman, et al., Nature (London, U.K.) 578, 237 (2020).

4. T. Antoja, A. Helmi, M. Romero-Gomez, et al., Nature (London, U.K.) 561, 360 (2018).

5. A. T. Bajkova and V. V. Bobylev, Astron. Lett. 38, 549 (2012).

6. L. N. Berdnikov, A. K. Dambis, and O. V. Vozyakova,
Astron. Astrophys. Suppl. 143, 211 (2000).

7. S. B. Bian, Y. Xu, J. J. Li, Y. W. Wu, B. Zhang,
X. Chen, Y. J. Li, Z. H. Lin, et al., Astron. J. 163, 54 (2022).

8. V. V. Bobylev, A. T. Bajkova, and A. S. Stepanishchev,
Astron. Lett. 34, 515 (2008).

9. V. V. Bobylev and A. T. Bajkova, Mon. Not. R. Astron. Soc. 408, 1788 (2010).

10. V. V. Bobylev and A. T. Bajkova, Mon. Not. R. Astron. Soc. 437, 1549 (2014).

11. V. V. Bobylev and A. T. Bajkova, Mon. Not. R. Astron. Soc. 447, L50 (2015).

12. V. V. Bobylev and A. T. Bajkova, Astron. Lett. 44, 676 (2018).

13. V. V. Bobylev, O. I. Krisanova, and A. T. Bajkova, Astron. Lett. 46, 439 (2020).

14. V. V. Bobylev and A. T. Bajkova, Astron. Rep. 65, 498 (2021).

15. V. V. Bobylev and A. T. Bajkova, Astron. Lett. 48 (2022, in press).

16. A. G. A. Brown, A. Vallenari, T. Prusti, et al. (Gaia Collab.), Astron. Astrophys. 649, 1 (2021).

17. J. Byl and M. W. Ovenden, Astrophys. J. 225, 496 (1978).

18. P. A. B. Galli, L. Loinard, G. N. Ortiz-L\'eon, M.
Kounkel, S. A. Dzib, A. J. Mioduszewski, L. F. Rodriguez,
L. Hartmann, et al., Astrophys. J. 859, 33 (2018).

19. T. Hirota, T. Nagayama, M. Honma, Y. Adachi, R. A. Burns, J. O. Chibueze, Y. K. Choi, K. Hachisuka, et al. (VERA Collab.), Publ. Astron. Soc. Jpn. 70, 51 (2020).

20. M. Honma, T. Nagayama, K. Ando, T. Bushimata, Y. K. Choi, T. Handa, T. Hirota, H. Imai, et al., Publ. Astron. Soc. Jpn. 64, 136 (2012).

21. K. Immer, J. Li, L. H. Quiroga Nu\~nez, M. J. Reid,
B. Zhang, L. Moscadelli, and K. L. J. Rygl, Astron. Astrophys. 632, A123 (2019).

22. C. C. Lin and F. H. Shu, Astrophys. J. 140, 646 (1964).

23. A. V. Loktin and M. E. Popova, Astrophys. Bull. 74, 270 (2019).

24. M. L\'opez-Corredoira, H. Abedi, F. Garz\'on, and
F. Figueras, Astron. Astrophys. 572, A101 (2014).

25. T. E. Lutz and D.H. Kelker, Publ. Astron. Soc. Pacif. 85, 573 (1973).

26. L. Martinez-Medina, A. P\'erez-Villegas, and A. Peimbert,
Mon. Not. R. Astron. Soc. 512, 1574 (2022).

27. A. M. Mel’nik, A. K. Dambis, and A. S. Rastorguev,
Astron. Lett. 27, 521 (2001).

28. Yu. N. Mishurov, I. A. Zenina, A. K. Dambis, A. M. Mel’nik, and A. S. Rastorguev, Astron. Astrophys. 323, 775 (1997).

29. Yu. N. Mishurov and I. A. Zenina, Astron. Astrophys. 341, 81 (1999).

30. I. I. Nikiforov and A. V. Veselova, Astron. Lett. 44, 81 (2018).

31. G. N. Ortiz-Le\'on, L. Loinard, S. A. Dzib, P. A. B.
Galli, M. Kounkel, A. J. Mioduszewski, L. F. Rodriguez,
R. M. Torres, et al., Astrophys. J. 865, 73 (2018).

32. G. N. Ortiz-Le\'on, K. M. Menten, T. Kaminski, A.
Brunthaler, M. J. Reid, and R. Tylenda, Astron. Astrophys.
638, 17 (2020).

33. A. S. Rastorguev, M. V. Zabolotskikh, A. K. Dambis,
N. D. Utkin, V. V. Bobylev, and A. T. Bajkova, Astrophys.
Bull. 72, 122 (2017).

34. M. J. Reid, K. M. Menten, X. W. Zheng, A. Brunthaler,
and Y. Xu, Astrophys. J. 705, 1548 (2009).

35. M. J. Reid, N. Dame, K. M. Menten, A. Brunthaler,
X.W. Zheng, Y. Xu, J. Li, N. Sakai, et al., Astrophys.
J. 885, 131 (2019).

36. N. Sakai, M. J. Reid, K. M. Menten, A. Brunthaler,
and T. M. Dame, Astrophys. J. 876, 30 (2019).

37. N. Sakai, T. Nagayama, H. Nakanishi, N. Koide, T. Kurayama, N. Izumi, T. Hirota, T. Yoshida, et al., Publ. Astron. Soc. Jpn. 72, 53 (2020).

38. N. Sakai, H. Nakanishi, K. Kurahara, D. Sakai, K.
Hachisuka, J.-S. Kim, and O. Kameya, Publ. Astron.
Soc. Jpn. 74, 209 (2022).

39. A. Sanna, M. J. Reid, T. M. Dame, K. M. Menten, and
A. Brunthaler, Science (Washington, DC, U. S.) 358,
227 (2017).

40. D. M. Skowron, J. Skowron, P. Mr\'oz, et al., Science
(Washington, DC, U. S.) 365, 478 (2019).

41. A. S. Stepanishchev and V. V. Bobylev, Astron. Lett. 39, 185 (2013).

42. L. Thulasidharan, E. D\'Onghia, E. Poggio, et al., arXiv: 2112.08390 (2021).

43. H.-F. Wang, M. L\'opez-Corredoira, Y. Huang, J. Chang, H.-W. Zhang, J. L. Carlin, et al., Astrophys. J. 897, 119 (2020).

44. L. M. Widrow, J. Barber, M. H. Chequers, and
E. Cheng, Mon. Not. R. Astron. Soc. 440, 1971 (2014).

45. Y. W. Wu, M. J. Reid, N. Sakai, T. M. Dame, K. M. Menten, A. Brunthaler, Y. Xu, J. J. Li, et al., Astrophys. J. 874, 13 (2019).

46. Y. Xu, S. B. Bian, M. J. Reid, J. J. Li, B. Zhang, Q. Z. Yan, T. M. Dame, K. M. Menten, et al., Astron. Astrophys. 616, L15 (2018).

47. Y. Xu, S. B. Bian, M. J. Reid, J. J. Li, K. M. Menten, T. M. Dame, B. Zhang, A. Brunthaler, et al., Astrophys. J. Suppl. Ser. 253, 9 (2021).

48. M. V. Zabolotskikh, A. S. Rastorguev, and
A. K. Dambis, Astron. Lett. 28, 454 (2002).
 }
\end{document}